# ExoPTF Science Uniquely Enabled by Far-IR Interferometry: Probing the Formation of Planetary Systems, and Finding and Characterizing Exoplanets

David Leisawitz[a], Tom Armstrong[b], Chad Bender[b,c], Dominic Benford[a], Daniella Calzetti[d], John Carpenter[e], William C. Danchi[a], Michel Fich[f], Dale Fixsen[a,g], Daniel Y Gezari[a], Matt Griffin[h], Martin Harwit[i], Alan J. Kogut[a], William D. Langer[j], Charles Lawrence[j], Dan Lester[k], Lee G. Mundy[g], Joan Najita[l], David Neufeld[m], Göran Pilbratt[n], Stephen Rinehart[a], Aki Roberge[a,o], Eugene Serabyn[j], Sachindev Shenoy[e,p], Hiroshi Shibai[q], Robert Silverberg[a], Johannes Staguhn[a,g], Mark R Swain[j], Stephen C. Unwin[j], Edward Wright[r], and Harold W. Yorke[j]

**Abstract**. *By providing sensitive sub-arcsecond images and integral field spectroscopy in the 25 – 400 μm wavelength range, a far-IR interferometer will revolutionize our understanding of planetary system formation, reveal otherwise-undetectable planets through the disk perturbations they induce, and spectroscopically probe the atmospheres of extrasolar giant planets in orbits typical of most of the planets in our solar system. The technical challenges associated with interferometry in the far-IR are greatly relaxed relative to those encountered at shorter wavelengths or when starlight nulling is required. A structurally connected far-IR interferometer with a maximum baseline length of 36 m can resolve the interesting spatial structures in nascent and developed exoplanetary systems and measure exozodiacal emission at a sensitivity level critical to TPF-I mission planning. The Space Infrared Interferometric Telescope was recommended in the <u>Community Plan for Far-IR/Submillimeter Space Astronomy</u>, studied as a Probe-class mission, and estimated to cost $800M. The scientific communities in Europe, Japan, and Canada have also demonstrated a keen interest in far-IR interferometry through mission planning workshops and technology research, suggesting the possibility of an international collaborative effort.*

This paper summarizes the "ExoPTF science" uniquely accessible to a far-IR spatial and spectral interferometer. Our recommendations are based on observations from the Spitzer Space Telescope and the Infrared Space Observatory (ISO), current technology, and the results of recently completed mission studies. Section 1 summarizes the scientific niche for a structurally-connected far-IR interferometer in the ExoPTF roadmap and the measurement capabilities required to achieve the science objectives. In §2, to the extent space permits, we offer mission implementation details based on a study of the Space Infrared Interferometric Telescope (SPIRIT), give a cost estimate and explain its provenance, and describe the key technology requirements. Programmatic issues relevant to the ExoPTF charter are discussed in §3.

The SPIRIT mission concept has deep roots in the astrophysics community's strategic plans. The 2000 Decadal Report Astronomy and Astrophysics in the New Millennium said: *"A rational coordinated program for space optical and infrared astronomy would build on the experience gained with [JWST] to construct SAFIR, and then ultimately, in the decade 2010 to 2020, build on the SAFIR, TPF, and SIM experience to assemble a space-based, far-infrared interferometer."* The kilometer baseline far-IR interferometer in the NASA Science Plan for astrophysics, widely known as the Submillimeter Probe of the Evolution of Cosmic Structure (SPECS), was the subject of a NASA "vision mission" study,[1] as was the Single Aperture Far-IR (SAFIR) Telescope.[2] In 2003, the *Community Plan for Far-IR/Submillimeter Space Astronomy*[3] expressed the consensus view of the far-IR community and offered practical advice; in addition to the flagship missions SAFIR and SPECS, the *Community Plan* called for two smaller missions, SPIRIT and a far-IR sky survey mission, identified key technologies, and recommended international collaboration. Unlike SPECS, SPIRIT does not require formation flying, and it was considered to be the logical way to embark on the interferometry path if its cost would be much less than that of a flagship mission. SPIRIT was selected for study as a candidate Origins Probe mission, its science potential was assessed, and its cost was estimated.[4] Because SPECS is beyond the time horizon of interest to the ExoPTF, only SPIRIT will be discussed in this white paper. Our recommendations are commensurate with the NASA budget for astrophysics, which leaves room for a mission like SPIRIT to enter Phase A in FY10.

---

[a] NASA GSFC; [b] Naval Research Lab; [c] NRC Research Associate; [d] UMass Amherst; [e] Caltech; [f] UWaterloo (Canada); [g] U Maryland; [h] Cardiff U (UK); [i] Cornell; [j] Jet Propulsion Laboratory, Caltech; [k] UT Austin; [l] NOAO; [m] Johns Hopkins U; [n] ESA; [o] NASA Postdoctoral Fellow; [p] IPAC; [q] Nagoya U (Japan); [r] UCLA

1 Science Goals and Measurement Requirements

In this section we explain how a particular detection technique – integral field spectroscopy in the far-IR – addresses the fundamental science issues of interest to the ExoPTF. Because SPIRIT employs this technique and adopts ExoPTF science for two of its three primary objectives, the most germane results from the SPIRIT Origins Probe mission study[4] are included here.

1.1 How Do Planetary Systems Form? How Do Habitable Conditions Develop?

The greatest obstacle to our understanding how planetary systems form is the unavailability of an observing tool that can decisively constrain theoretical models. The early phase of planet formation takes place behind a veil of dust, rendering protoplanetary disks inaccessible to visible wavelength telescopes. Further, until now, the far-infrared light from protoplanetary disks, which is their dominant emission, has only been seen as an unresolved blur. **Only spatially-resolved far-IR observations of protoplanetary disks have the power to break model degeneracy and teach us how planetary systems form.** Sub-arcsecond angular resolution will be needed to resolve the structures of interest. To study the early gas-rich phase, when giant planets form, we will need to resolve objects in regions like ρ Oph and Taurus, at 140 pc. The gas-poor terrestrial planet formation phase can be observed in the Tuc or TW Hya associations whose distances are ~50 pc. The Spitzer Space Telescope and the planned observatories Herschel and SOFIA (the Stratospheric Observatory for Infrared Astronomy) lack the necessary angular resolution by more than an order of magnitude in the far-IR (Figure 1).

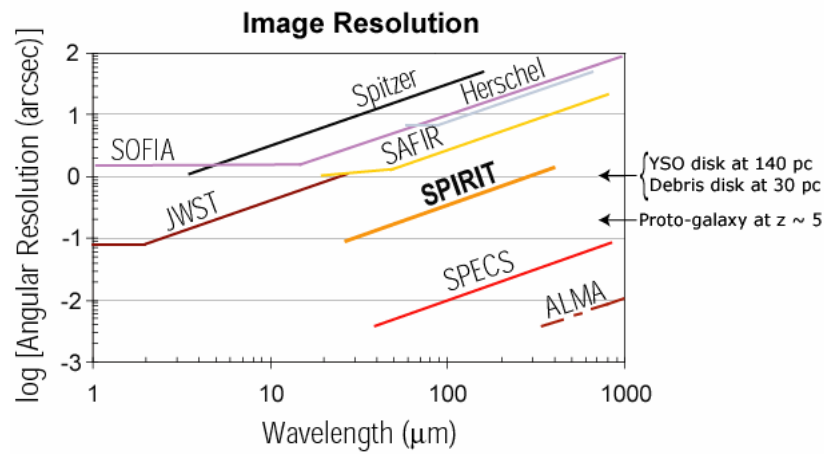

**Fig. 1** – Angular resolution available with current, next-generation, and future infrared and submillimeter observatories. SPIRIT will provide one hundred times better angular resolution than the Spitzer Space Telescope. The resolution is comparable to that of JWST, but at ten times longer wavelengths, where Young Stellar Objects (YSOs) emit most of their light.

To gain real astrophysical insight into the planet formation process, in addition to resolving protoplanetary disks spatially, we will have to exploit information available in the spectral domain (Figure 2). The far-IR continuum emission is dominated by thermal emission from dust. A spatial map of the far-IR spectral energy distribution will enable future observers to measure the three-dimensional distribution of solid state material and probe the dust temperature locally. **By measuring the gas contents of planet forming disks of various ages it will be possible to constrain the timescale for gas giant planet formation and the migration of planetary bodies of all sizes.** Direct observations of $H_2$ and HD in the readily excited 28 μm and 112 μm rotational lines will lessen our dependence on surrogate gas tracers, such as CO, which can be photodissociated or frozen onto grain surfaces. Numerous far-IR lines from hydrides, such as CH and OH, will be measured with Herschel and mapped with SPIRIT, as will the strong [C I], [O I] and [C II] fine structure lines at 370, 146, 63 and 158 μm. When coupled with models,[5,6,7] far-IR spectral line observations will give us tremendous insight into the chemistry and physical conditions in young planet forming disks. The C/O ratio, derivable from these observations, is thought to affect the composition, surface chemistry, and perhaps the habitability of planets.[8,9]

Because of its biological significance, water will be one of the most interesting molecules to observe. Most of the volatiles found on the terrestrial planet surfaces are thought to be delivered by impacts of small $H_2O$ rich bodies from the outer solar system. ALMA will probably not be able to detect water vapor through the atmosphere, or will detect it rather noisily. **A space-based interferometer can access the rich far-IR spectrum of the $H_2O$ molecule (Figure 2c, in red), map the water distribution in young protoplanetary disks to study the formation of the water reservoir, and search for evaporating water in extrasolar comet trails.**[10, 11] **The interferometer will make complementary observations of frozen water by mapping the $H_2O$ ice features at 44 and 63 μm.**[12]

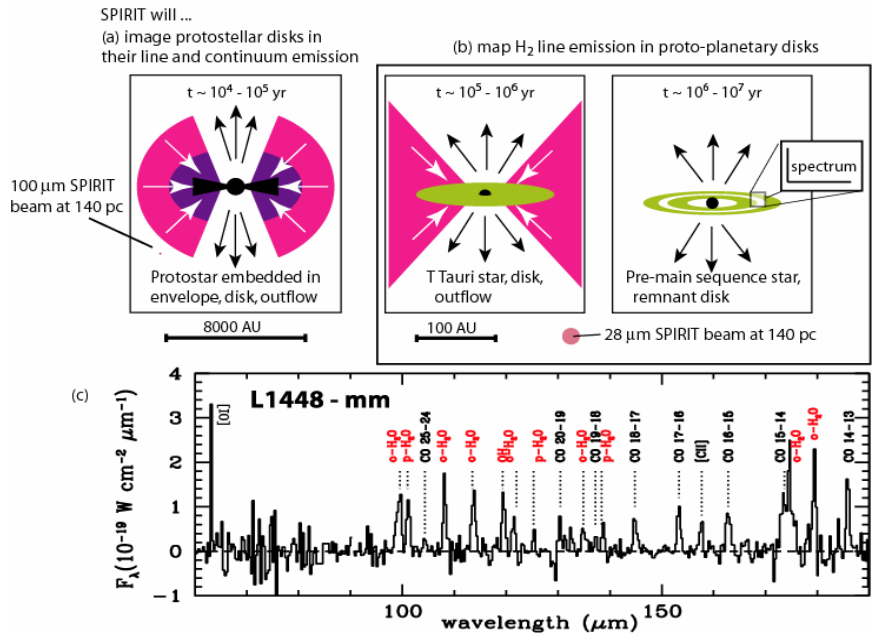

**Fig. 2** – SPIRIT, which naturally produces spatial-spectral "data cubes," will (a) image protostellar disks in their line and continuum emission, and (b) map $H_2$, $H_2O$, and C and O fine structure line emission in protoplanetary disks, providing a spectrum like the one shown in panel (c) at each of many spatial locations. The continuum-subtracted spectrum (c) is based on ISO observations of a Class 0 protostar.[13]

Spitzer, renowned for its observing capabilities, is a useful sensitivity benchmark. SPIRIT will have cold (~4 K) mirrors like Spitzer, light collecting area 3.2x that of Spitzer, and more advanced detector technology. Our instrument model[4] predicts SPIRIT line sensitivity comparable to that of the IRS instrument on Spitzer and the PACS instrument on Herschel (bigger telescope, but warmer). The SPIRIT spectral resolution will be an order of magnitude greater than that shown in Fig. 2c. All of the spectral lines seen in Figure 2c will be detectable, even if the line flux is divided evenly among more than 300 spatial resolution elements.

1.2 Dust Concentrations in Debris Disks: Signposts of Exoplanets

**Spatially resolved far-IR images of debris disks can reveal hidden planets in a manner complementary to the motion-dependent (astrometric or Doppler) techniques.** The orbits followed by dust grains in developing and established planetary systems (i.e., debris disks) are

perturbed gravitationally by the planets. Orbital resonances can produce dust concentrations,[14,15] which have been observed at visible to millimeter wavelengths.[16,17,18,19] This dust glows most brightly in the spectral range ~20 – 300 μm, where main sequence stars are faint. The locations, masses and orbits of unseen planets can be deduced from the shapes and temporal variations in the dust debris disks,[14,15] just as new Saturnian moons have been found after ring gaps and features divulged their hiding places (Fig. 3).

Although the nearest debris disks are only a few parsecs away and resolved by Spitzer (Fig 3a), an important objective is to understand our own solar system in the context of a representative sample of exoplanetary systems. At a minimum, it will be necessary to detect 1 AU structures in debris disks out to a distance of ~10 pc, implying an angular resolution requirement of 0.1 arcsec. At 25 μm, JWST will miss that target by an order of magnitude (Figure 1), but SPIRIT will be able to image 19 known luminous debris disks[20] and many more faint disks. **Vital to planning TPF-I, SPIRIT has the sensitivity to detect a 3 "zodi" disk (1 zodi = $10^{-6}L_{Sun}$) or map the distribution of dust in a resolved 100 zodi disk at 10 pc in a 1-week observation period.**

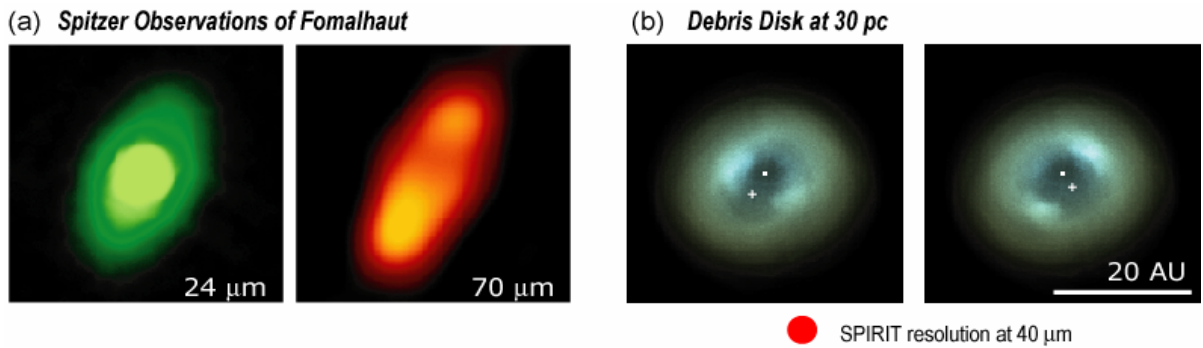

**Fig. 3** –Spitzer resolves four nearby debris disks, including Fomalhaut, shown here (a) at 24 and 70 μm.[17] With angular resolution a hundred-fold better than that of Spitzer, SPIRIT will provide clear images of a large statistical sample of debris disks, enabling discoveries of new planets and a great improvement in our understanding of the factors that influence the evolution of planetary systems. The model images in (b), based on Eps Eri but scaled to 30 pc, show the predicted far-IR emission at 40, 60, and 100 μm color-coded as blue, green, and red, respectively. The dust-trapping planet (+) is shown at two orbital phases, and the resonantly trapped dust grains can be seen to have moved.

1.3 Probing the Atmospheres of Gas Giant and Ice Giant Exoplanets

Recent Spitzer observations of a transiting extrasolar giant planet demonstrate the value of IR spectroscopy as a tool to constrain a planet's temperature structure and probe the composition of its atmosphere.[21] In addition to exploiting the transiting technique, which favors detections of "hot" exoplanets, a far-IR interferometer equipped with a scanning optical delay line can measure the spectra of giant exoplanets at larger orbital distances. An extrasolar planet's light would appear as the modulating signal component in a multi-epoch series of interferograms, obtained with the planet at different orbital positions, while starlight would produce a stable fringe pattern. With sufficient signal-to-noise ratio and a simple astrometric model, the planet's interferogram could be extracted and Fourier transformed to obtain the desired spectrum, and the orbital parameters could be measured. The spectral resolution will be dictated by the scan range of the delay line and could be $\lambda/\Delta\lambda \sim 3000$, or some of the spectral resolution could be sacrificed to improve the signal-to-noise ratio. Taking Jupiter as an example,[22] we can expect to detect broad $NH_3$ bands, which dominate the spectrum from 40 to 100 μm, and study the abundances of key chemical species such as water and methane. These measurements will be challenging, but not more difficult than the differential measurements (star + planet minus star) required in the transiting technique. Starlight nulling is not required if the exoplanet is separated by more than 0.1 arcsec (1 AU at 10 pc) from the star. **SPIRIT will be capable of measuring the far-IR spectra of extrasolar giant planets in close orbits during transits, and in orbits characteristic of most of the planets in our solar system, helping us to understand the internal structure, formation, and migration of these planets to smaller orbits, and to measure the chemical abundances of their atmospheres.**

1.4 Summary of Measurement Requirements

During the Origins Probe mission study the science goals discussed in §§1.1 and 1.2 were given top billing, along with a third goal (to learn how high-redshift galaxies formed and merged to form the present-day population of galaxies), and corresponding measurement requirements were derived and used as the basis for an engineering design and cost estimate.[4] Table 1 summarizes the SPIRIT design requirements, most of which were mentioned above. A typical SPIRIT observation will take about 1 day and yield, after ground data processing, a "data cube" with two high-resolution spatial dimensions and a third spectral dimension.

**Table 1**. SPIRIT Measurement Requirements

| | |
|---|---|
| Wavelength range | 25 to 400 μm |
| Instantaneous field of view | 1 arcmin |
| Angular resolution | 0.3 (λ/100 μm) arcsec |
| Spectral resolution (λ/Δλ) | 3000 |
| Point source sensitivity | $3 \times 10^{-18}$ W m$^{-2}$ spectral line, 5σ in 1 hr<br>~1 μJy continuum in "deep field" |
| Typical time per target field | 29 hrs |
| Field of regard | 40° band centered on ecliptic plane |
| Mission life, on station | 3 years (propellant for 5) |

2  Mission Concept and Enabling Technology

A single instrument will provide the required measurement capabilities. SPIRIT has two 1-m diameter light-collecting telescopes cryo-cooled to 4 K and a central beam-combiner attached to a deployable structure (Figure 4). The telescopes are mounted on trolleys, which move along rails to provide interferometric baselines ranging in length from 6 m to 36 m. The interferometer rotates during an observation with the rotation axis pointing toward the target field. The "u-v plane" sampling can be tailored to the expected spatial brightness structure in the scene and can be dense, so SPIRIT will produce very good images. For each baseline observed, an optical delay line is scanned, yielding a set of white light interferograms, one for each pixel in SPIRIT's four pairs of detector arrays. Like its laboratory prototype, the Wide-field Imaging Interferometry Testbed,[23,24,25] SPIRIT provides a spectrum in every spatial resolution element; the delay line is scanned through the distance required to (a) equalize path lengths through the two arms of the interferometer for all angles in a 1 arcmin FOV, and (b) provide spectral resolution R = 3000.

The most critical subsystems (optical, cryo-thermal, detector, metrology, and mechanisms) and the SPIRIT operations concept are described in Reference [4], but all of the essential elements of mission design were addressed during the SPIRIT study, including, for example, the attitude control system, power system, and flight software. The August 2007 SPIE Conference "UV/Optical/IR Space Telescopes: Innovative Technologies and Concepts" will feature a series of papers on the SPIRIT overall engineering design,[26] optical design,[27] cryo-thermal design,[28] mechanical design and mechanisms,[29] and detectors.[30]

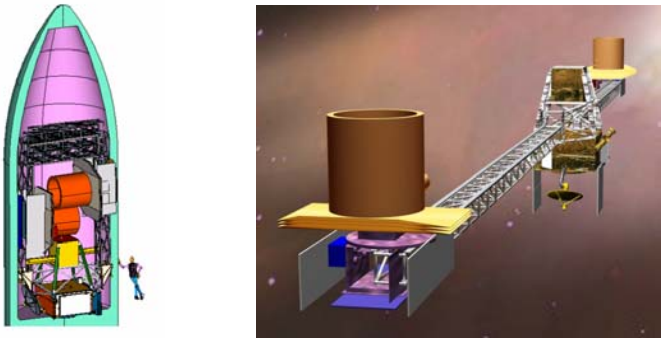

**Fig. 4** – SPIRIT and its expendable launch support structure, when stowed for launch (left), are 8.7 m tall and were designed to fit into an Atlas V 5-medium fairing. The deployed observatory is shown in an artist's concept (right).

All of the key enabling technologies for SPIRIT are maturing, but **we recommend investment in the most challenging areas: far-IR detectors with sensitivity $10^{-19}$ W Hz$^{-1/2}$, speeds of ~150 µs, and pixel counts of order 100; experimental verification of a cryo-thermal system model; and performance verification of the algorithms used to produce spatial-spectral data cubes from interferometric data.**

2.1 Cost Estimate

The cost of SPIRIT was estimated by the SPIRIT Origins Probe mission study team based on the "grass-roots" analysis of engineers and scientists with experience in space flight projects and space flight hardware development. Each element of the payload cost was accompanied by a "Basis of Estimate" and a description of the Work Breakdown. A 72 month development period from Phase B start to Launch was assumed. The spacecraft bus cost was estimated based on extrapolation from a comparable bus whose cost had been estimated earlier for a different mission study. Standard contingencies, ground support equipment, integration and test, and launch costs were included. The cost for technology development was not included. The estimated total mission cost in constant $ FY04 was $800M.

To check the grass roots cost estimate an independent cost model was developed using PRICE-H. Only the spacecraft and payload costs were modeled. The PRICE-H model agreed with the team's estimated spacecraft plus payload costs to within about 10%.

3 Programmatic Issues

**In addition to enabling tremendous gains on the exoplanetary science frontier, a structurally connected far-IR interferometer costing less than $1B is a very practical place to start on the path toward progressively more complex space interferometry missions.** Interferometric space missions, such as SIM and TPF-I/Darwin, will ultimately be needed to tackle many of the ExoPTF's science goals. Interferometry is all about wavefront control, which grows progressively challenging as the observed wavelength decreases. Further complications arise when starlight nulling is required, or when the interferometric baseline is so large that a structure to hold the light collecting telescopes becomes impractical. Far-IR spatial-spectral interferometry is largely immune to the technical risks that will have to be confronted to launch successful visible wavelength or long baseline mid-IR nulling interferometers.

**Finally, we recommend the opening of a dialog between NASA and its partners in ESA, CSA, and JAXA to explore the possibility of international collaboration on a space-based**

**far-IR interferometry mission.** Favorable conditions exist for international collaboration on such a mission. The European far-IR community has its sights set on the post-Herschel mission era and ESA's recently-released *Cosmic Visions* call for proposals. Three European community workshops, in 2003, 2005, and 2006, culminated in the conclusion that a far-IR long-baseline interferometer is the logical successor to Herschel. The ESA Concurrent Design Facility published a far-IR connected interferometry mission study report in 2006.[31] The Canadian space astrophysics community held an informal workshop in November 2006, organized by the CSA, and a far-IR interferometer was the only project clearly identified as the highest priority by the majority of the participants. The first flight of the Japanese balloon-borne Far-Infrared Interferometric Telescope Experiment is scheduled in the Fall of 2007 from Brazil. Although international collaboration on a space mission is challenging, the modularity inherent in the design of an interferometer can help to offset some of the barriers, as can an early partnership agreement.